\documentclass[usenatbib]{mn2e}

\usepackage{amssymb}
\usepackage{epsfig}
\usepackage{colordvi}
\usepackage{txfonts}
\usepackage{graphicx}

\newcommand{\bez}{\begin{eqnarray*}}
\newcommand{\eez}{\end{eqnarray*}}
\newcommand{\be}{\begin{equation}}
\newcommand{\ee}{\end{equation}}
\newcommand{\beq}{\begin{eqnarray}}
\newcommand{\eeq}{\end{eqnarray}}
\newcommand{\bc}{\begin{center}}
\newcommand{\ec}{\end{center}}

\def\msun{{\rm M_{\odot}}}

\def \mum{\mu{\rm m}}
\def \fd{f_{\rm d}}
\def \Nh{n_{\rm H_2}}
\def \fw{f_{\rm H_2O}}
\def\Rout{R_{\rm out}}
\def\Rin{R_{\rm in}}

\title[A warped $m=2$ water maser disc in V778 Cyg?]
{A warped $m=2$ water maser disc in V778 Cyg?}

\author[N. Babkovskaia et al.]
{Natalia Babkovskaia,$^{1}$\thanks{E-mail:
natalia.babkovskaia@oulu.fi (NB) }
Juri Poutanen,$^1$
Anita~M.~S. Richards,$^{2}$  and Ryszard Szczerba$^3$    \\
$^1$Astronomy Division, PO Box 3000, FIN-90014 University of Oulu, Finland\\
$^2$Jodrell Bank Observatory, University of Manchester, Macclesfield,
    Cheshire SK11 9Dl \\
$^3$N. Copernicus Astronomical Center, Rabia\'nska 8, 87-100 Toru\'n, Poland
}

\pagerange{\pageref{firstpage}--\pageref{lastpage}}
\begin{document}

\maketitle
\label{firstpage}

\begin{abstract}
The silicate carbon star V778 Cyg is a source of 22 GHz water maser
emission which was recently resolved by MERLIN.  Observations revealed
an elongated $\cal{S}$-like structure along which the velocities of
the maser features show a linear dependence on the impact
parameter. This is consistent with a doubly-warped $m=2$ disc observed
edge-on.  Water masers and silicate dust emission (detected by {\it
IRAS} and {\it ISO}) have a common origin in O-rich material and are
likely to be co-located in the disc. We propose a detailed
self-consistent model of a masing gas-dust disc around a companion
to the carbon star in a binary system, which allows us to estimate the
companion mass of $1.7 \pm 0.1$ M$_{\odot}$, the disc radius of $40\pm
3$ AU and the distance between companions of $\sim 80$ AU.  Using a
dust-gas coupling model for water masing, we calculate the maser power
self-consistently, accounting for both the gas and the dust energy
balances.  Comparing the simulation results with the observational
data, we deduce the main physical parameters of the masing disc, such
as the gas and dust temperatures and their densities. We also present
an analysis of the stability of the disc.
\end{abstract}



\section{Introduction}

Carbon stars  with amorphous silicate dust features at 10 and 18 $\mum$
(silicate carbon stars) which typify O-rich dust envelopes  \citep{LM86},
were first discovered with {\it
the Infrared Astronomical satellite (IRAS)}.
A distinguishing feature of this class of carbon stars is a mixed
chemistry, i.e. simultaneous presence of O- and C-rich material
around them. The {\it Infrared Space Observatory (ISO)} revealed
that this mixed chemistry phenomenon is not unique.  About 20
silicate carbon stars are known in our Galaxy \citep{CW99a,Sz02}.

The best known example of this class is V778 Cyg. Infrared
observations  show that the shape and intensity of the silicate
features in the spectrum of V778 Cyg did not change during a period of about
14 years \citep[][ hereafter Y00]{YD00}. This indicates that
the O-rich material is located  in some stable configuration. Y00
proposed that it is stored in a disc around an invisible (most probably
main-sequence) star, which is the companion of the carbon star  in a
binary system \citep[see also][]{MG87,LE90,BK91}. V778 Cyg also shows
$\rm H_2O$ and OH maser emission. Single dish monitoring of water maser
emission during 12 years shows a very small velocity shift ($<0.5$
km s$^{-1}$) of  the main components in the  maser spectrum
\citep{EL94}\footnote{see also  http://www.hs.uni-hamburg.de/DE/Ins/Per/ \\
Engels/engels/wcatalog.html}.

In October 2001 water maser emission from V778 Cyg was imaged
using five telescopes of MERLIN \citep{SS06}.
The position of the maser complex was found to be displaced
by $\sim$190\,mas from the position of the C-star measured 10 years previously,
which is unlikely to be a result  of the proper motion. Instead, this can be
considered as an additional evidence that the O-rich material is stored
around the secondary \citep{SS06}.

The MERLIN data show that the masers are distributed in an $\cal{S}$-shaped figure.
In this paper, we construct a model of the spatial and
velocity distribution of the water masers around V778 Cyg. We
develop a self-consistent description of the physical mechanism of
the water maser pumping together with an explanation for the
observed infrared silicate dust emission.  We also determine the
main parameters of the system.

We present the main observational data in Section~\ref{sec:data}.
We model the 22-GHz water maser spatial and velocity distributions and
the infrared dust emission as well as the physical mechanism of maser
pumping in Section~\ref{sec:modelling}.
We discuss our results in Section~\ref{sec:discussion} and
conclude in Section~\ref{sec:conclusion}.

\begin{figure*}
  \begin{center}
    \begin{minipage}[c]{11cm}
      \epsfig{file=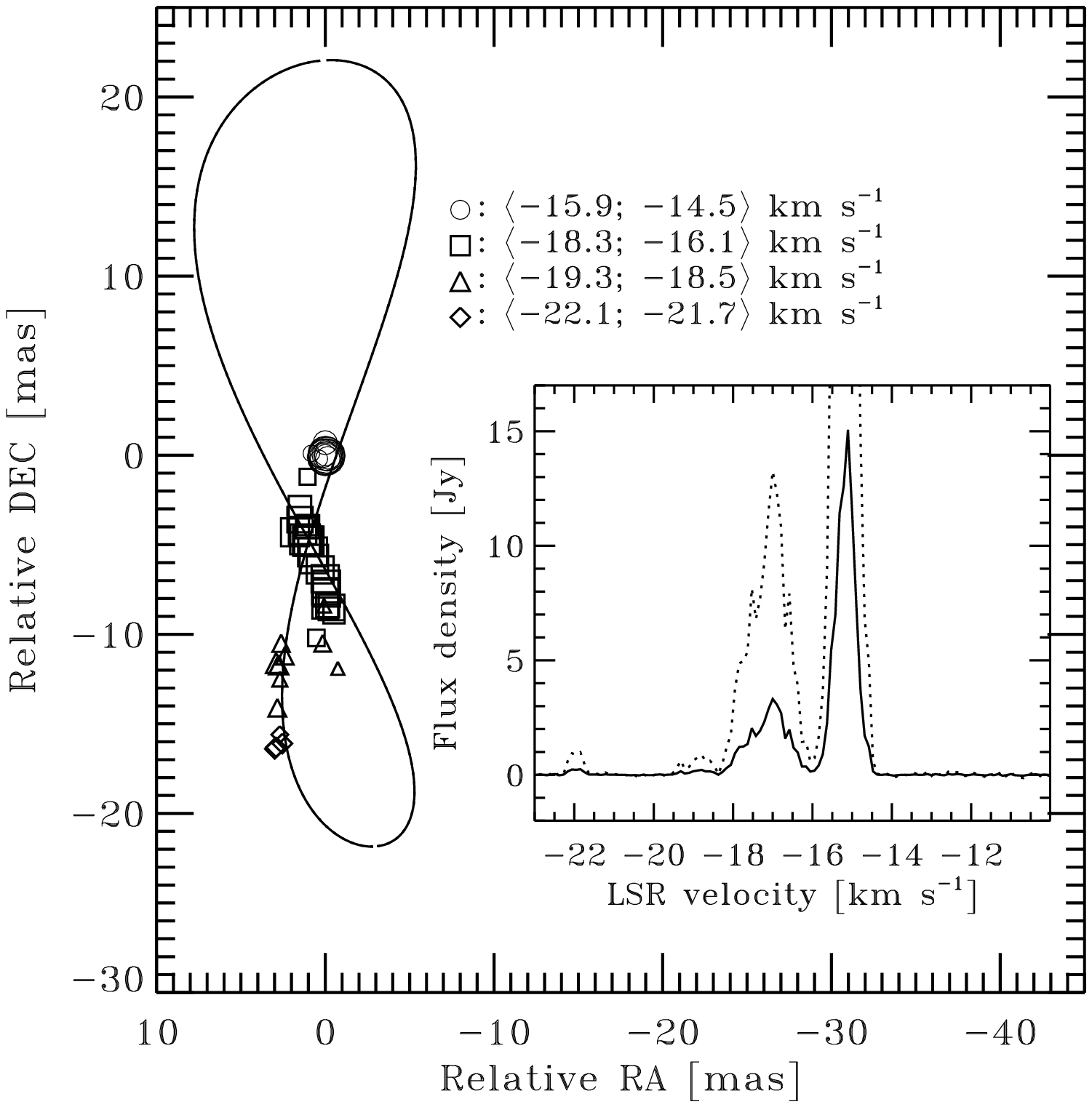,width=\textwidth}
    \end{minipage}\hfill
    \begin{minipage}[c]{5cm}
      \caption{Water maser distribution observed towards V778 Cyg \citep{SS06}.
The coordinates are measured relative to the position of the strongest
feature at $-15$ km s$^{-1}$ at
$(\alpha, \delta) = (20^{\rm h}36^{\rm m}07^{\rm
s}.38327$, $60\degr05\arcmin26\farcs024)$.
The solid curve shows the best fit to a model of a doubly-warped
ring of radius 40 AU (at the assumed distance of 1.8 kpc) which is thin in the radial direction;
 other parameters are presented in Table~1.
The ring centre is at the origin of the coordinate system.
The insert shows the MERLIN spectrum of the
$\rm H_2O$ maser emission. The dotted curve shows the spectrum
magnified by a factor of four to enlarge the weak features.
      \label{fig:geom}}
    \end{minipage} \hspace*{0.5cm}
  \end{center}
\end{figure*}

\section{Data}
\label{sec:data}

\begin{figure}
\centerline{ \epsfig{file=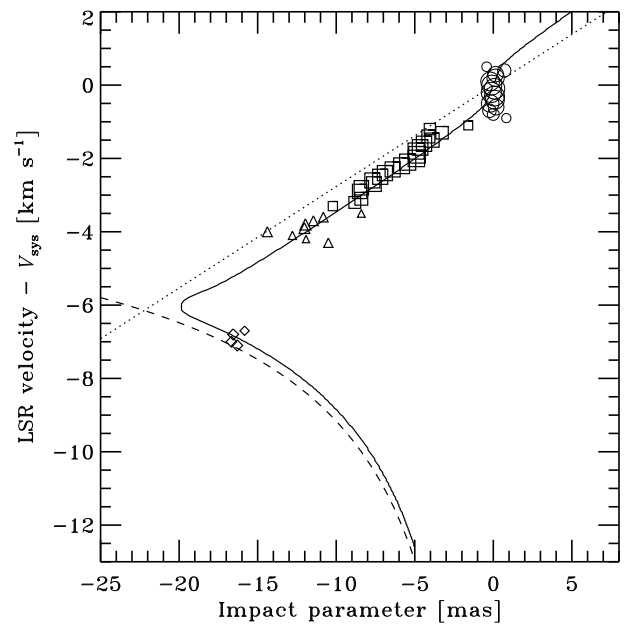,width=8.5 cm}}
\caption{Dependence of the velocities of the maser features on the
impact parameter measured along the major axis of the observed
structure, relative to the brightest feature at $-15$ km s$^{-1}$,
which is assumed to be the systemic velocity $V_{\rm sys}$. The
solid curve represents the model velocity dependence corresponding
to a Keplerian disc with the inner and outer  radii of $\Rin=10$
and $\Rout=40$~AU, with $V(b)$ computed from $\partial \tau_{\rm
m}/\partial b=0$, where $\tau_{\rm m}$ is given by
Eq.~(\ref{eq:tau}).
The linear dependence $V(b)=V_{\rm K}(\Rout) b/\Rout$ is shown by
a dotted line for comparison. The pure Keplerian rotation law
$V(b)\propto 1/\sqrt{b}$ is shown by a dashed curve. }
\label{fig:vel}
\end{figure}

The silicate carbon star V778 Cyg is located at  a distance
of about $d=1.8$ kpc, while
the carbon star luminosity is $L_{*} \simeq 5.6 \  10^3 {\rm L_{\odot}}$
\citep{BL96}.
Observations of V778 Cyg at 22 GHz by MERLIN revealed that the
water maser features form an $\cal{S}$-shaped structure with an angular
size of 18 mas (\citealt{SS06}; see Fig.~\ref{fig:geom}).  The
observed line-of-sight velocities of masing blobs depend almost
linearly on the impact parameter measured along the major axis of the
structure as shown in Fig.~\ref{fig:vel}.  The total spectrum of water
maser emission (see insert in Fig.~\ref{fig:geom}) consists of four
major components near $-22$, $-19$, $-17$ and $-15$ km s$^{-1}$.  The
last component is the strongest, with a central peak of about 15 Jy,
which corresponds to a lower limit to the brightness temperature of
$T_{\rm b}=6 \ 10^8$ K \citep{SS06}.

\section{Modelling}

\label{sec:modelling}

\begin{figure*}
\begin{center}
\leavevmode \epsfxsize=8cm \epsfbox{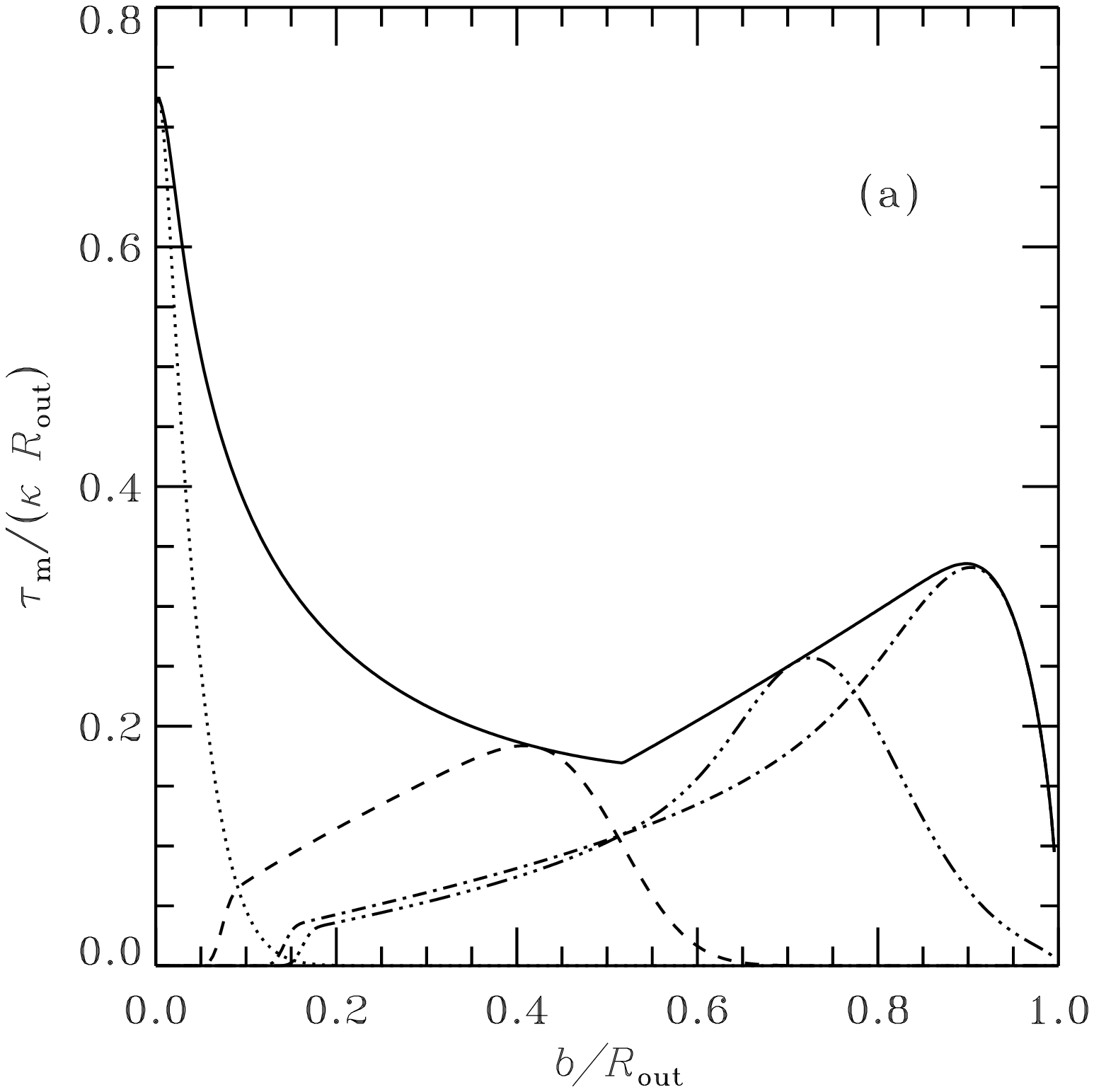} \hspace{1cm} \epsfxsize=8cm \epsfbox{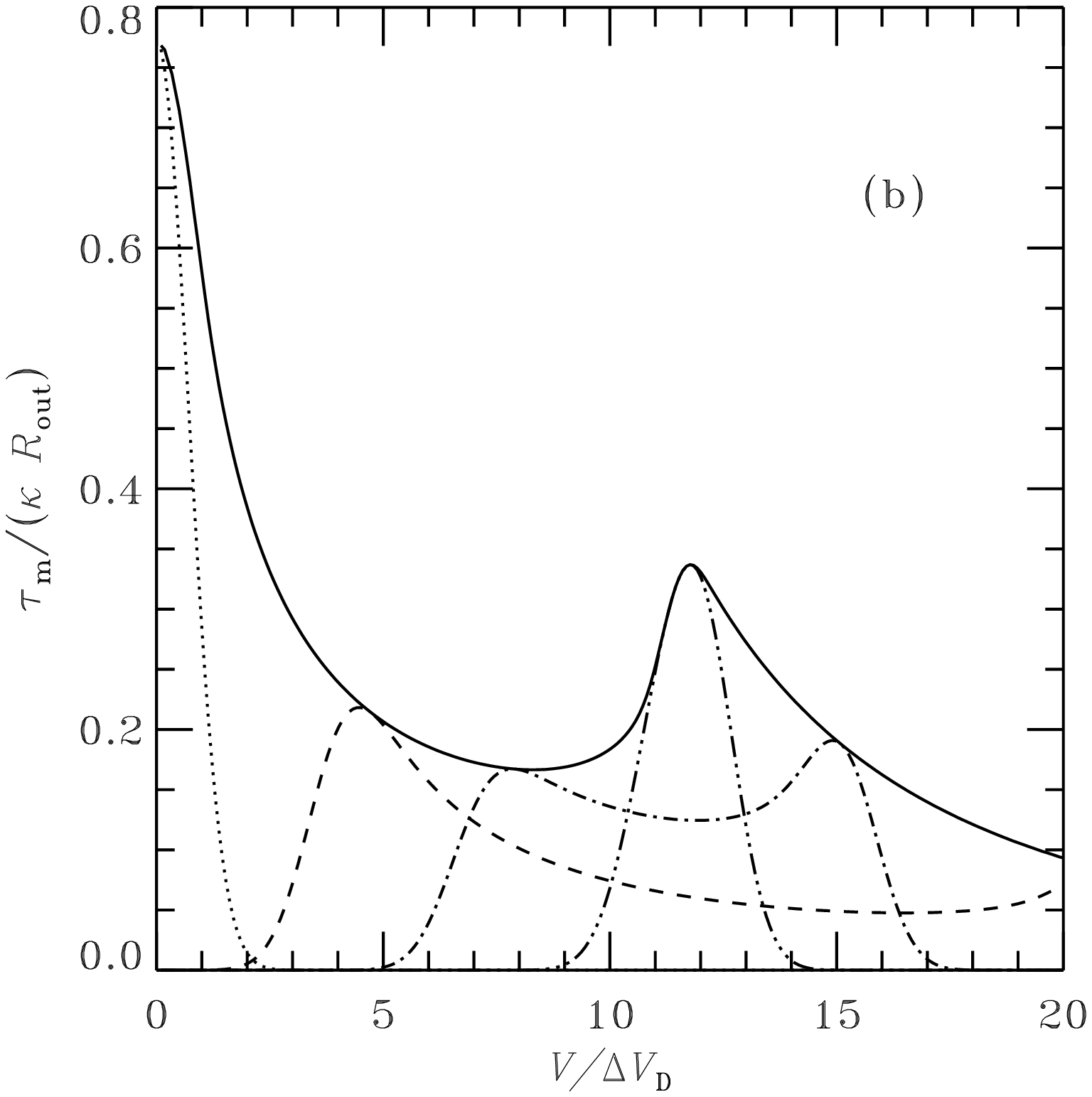}
\end{center}
\caption { The optical depth in a  masing line $\tau_{\rm m}$ (in
units $\kappa \Rout$) for $\xi=V_{\rm K}(\Rout)/\Delta V_{\rm
D}=11.7$ and the ratio of the inner-to-outer disc radii $\Rin/\Rout=1/4$.
The parameters are chosen to represent a disc around a $1.7 \msun$ star,
with the outer radius $\Rout=40$ AU and the gas temperature of 300 K.
(a)~Dependence on the impact parameter $b$ for the velocities
$V'=V/\Delta V_{\rm D} =0$ ({\it dotted curve}), 5.7 ({\it dashed
curve}), 11.4  ({\it dot-dashed curve}), 13.3 ({\it triple
dot-dashed curve}). (b)~Dependence  on the velocity of maser
features for $b/\Rout= 0$ ({\it dotted curve}), $0.3$ ({\it dashed
curve}), $0.6$ ({\it dot-dashed curve}), $0.9$ ({\it triple
dot-dashed curve}). Solid curves show the maximum $\tau_{\rm m}$
as a function of $b$ and $V$ (see Section \ref{sec:Kep_disk}).}
\label{fig:tau_lv}
\end{figure*}

\citet{SS06} show that the structure and velocity field of the
maser features (presented in Figs.~\ref{fig:geom} and
\ref{fig:vel}) are consistent with a Keplerian disc, observed
almost edge-on, and present a simple quantitative interpretation
of the observational data. The three maser spectral components  at
$-19$, $-17$, and $-15$ km s$^{-1}$ (see Fig.~\ref{fig:geom}) have
been detected in every observation during the whole period from
1988 to 2002. They are symmetric around $-17$ km s$^{-1}$ in
velocity space. They are also almost equally displaced on the sky.
By analogy with water masers in AGNs, this has prompted the
interpretation that these are the systemic and the high velocity
components that appear towards the centre and in the wings of a
Keplerian disc \citep[see e.g.][]{MM95,BG00}. In that case the
velocities of the individual features making up the systemic
component should show a linear dependence on the impact parameter
$V(b)\propto b$ \citep{GG83,WW00} which indeed is the case (see
Fig.~\ref{fig:vel}).  This linear dependence could be interpreted
as solid-body rotation. However, such an explanation is highly
unlikely, because this requires a self-gravitating disc having a
mass comparable to or even larger than the central mass. In such a
case, the  number density in the disc would be too high for the
maser to operate as the population levels of water molecules would
become thermalised by collisions with hydrogen.

On the other hand, the Keplerian disc interpretation implies a
velocity at the outer radius of $V_{\rm K}\approx2$~km~s$^{-1}$
and a disc radius of $R \simeq 11 $ AU.  The component at
$-22$~km~s$^{-1}$, which appeared in the spectrum in 1992, causes
a problem for this model.
 It lies farthest from the disc centre but  has too large a velocity
 for a simple Keplerian disc.

The $\cal{S}$-shape is likely to result from
the curving  of the disc \citep{SS06}.
However, the observed displacement of maser
features at $-19$~km~s$^{-1}$ relative to those at $-17\pm1$ km s$^{-1}$ is very
large and is comparable to the estimated disc radius, which would
be inconsistent with a simple model of a {\it slightly warped}
masing disc, widely accepted, for example, for NGC 4258
\citep[e.g.][]{MM95}. The disc radius thus could be larger than the assumed
11~AU (corresponding to 6 mas), and the maser emission at
$-22$~km~s$^{-1}$ could still   come from the disc.

The stability of the dust spectrum (Y00) and of the water maser
spectral features \citep{EL94} prompts us to suggest that they
have a common origin in the O-rich material in the disc.
This disc is unlikely to be located
around the carbon star, because the dust would be blown away by the
stellar radiation at the distance corresponding to the observed
disc size.
It is thus natural to accept the hypothesis of Y00 and
consider a disc around the companion of the carbon star in a binary system.

\subsection{Masers from a Keplerian disc}

\label{sec:Kep_disk}

Let us first consider a flat thin disc in Keplerian motion.  The
theory of masers from Keplerian discs is well developed
\citep{GG83,WW00}.  Masers are bright when the line of sight crosses
the disc at a grazing angle, close to edge-on.  Due to the velocity
gradient, the maximum maser coherence length is reached towards the
disc centre (where the matter moves perpendicular to the line of
sight) and near the outer parts (where the gas motion is almost
parallel to the line of sight).  The strongest spectral features thus
appear at the systemic velocity and at  velocities shifted towards
the red and blue by almost the Keplerian velocity near the outer
edge of the masing disc.

The optical depth in a masing line, integrated along the line of
sight $s$, as a function of the impact parameter $b$ and velocity
$V$ is given by \citep[see, for example,][]{GG83,WW00} \be
\label{eq:tau_kepler} \tau_{\rm m}(b,V)= \int \kappa\ \exp
\left\{-\left(\frac{ b \ \Omega_{\rm K}(r)- V}{ \Delta V_{\rm D}}
\right)^2\right\} {\rm d} s, \label{eq:tau} \ee where $\kappa$ is
the maser opacity, $\Omega_{\rm K}(r)=V_{\rm K}(r) /r$ is the
Keplerian angular velocity at radius $r=\sqrt{s^2+b^2}$, $\Delta
V_{\rm D} =(2 k T/m_{\rm H_2O})^{1/2}$ is the thermal velocity for
water molecules, and $T$ is the gas temperature.  The upper
integration limit $\sqrt{\Rout^2-b^2}$ is determined by the outer
disc radius $\Rout$. The lower limit is either zero or
$\sqrt{\Rin^2-b^2}$ (corresponding to the inner disc radius
$\Rin$). Equation (\ref{eq:tau_kepler}) gives half of the total
optical depth along the line of sight.  The optical depth can be
expressed as a function of dimensionless variables $b'=b/\Rout$
and $V'= V/\Delta V_{\rm D}$: \be \label{eq:tauint}
\frac{\tau_{\rm m}(b',V')}{\ \Rout} = \int \kappa \ \exp
\left\{-\left( \frac{\xi\ b'}{r'^{\ 3/2}}- V' \right)^2\right\}
{\rm d} s' , \ee where $r'$ and $s'$ are the corresponding
distances measured in units of $\Rout$, and $\xi=V_{\rm K}
(\Rout)/\Delta V_{\rm D}$ is the ratio of the Keplerian velocity
at the outer disc edge to the thermal velocity. At constant
opacity $\kappa$ and temperature $T$, the integral in
Eq.~(\ref{eq:tauint}) depends only on $\xi$ and on the disc
geometry expressed through the ratio of the inner to the outer
disc radii $\Rin/\Rout$.

\begin{figure*}
\begin{center}
\leavevmode \epsfxsize=8cm \epsfbox{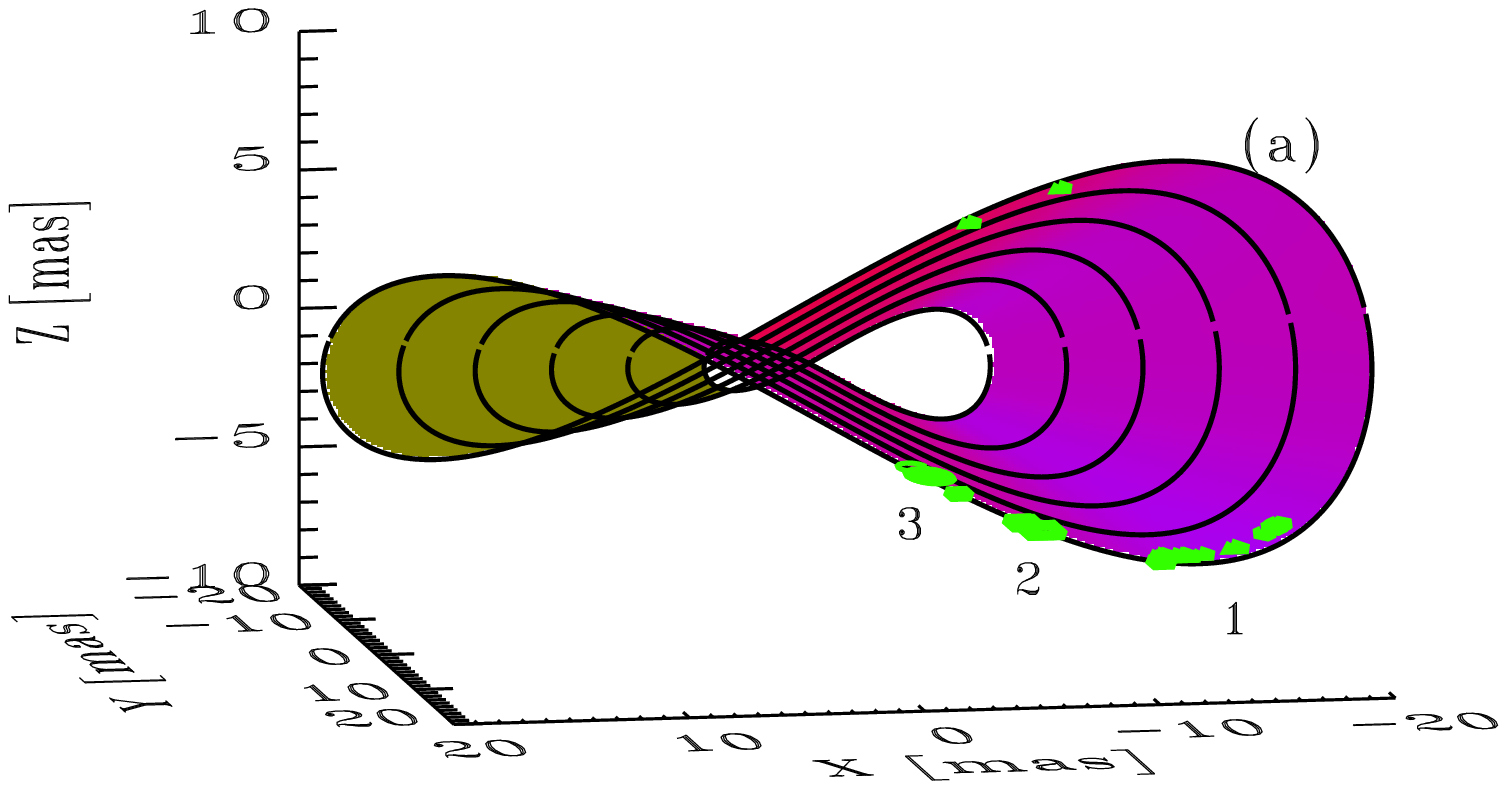} \epsfxsize=9cm
\epsfbox{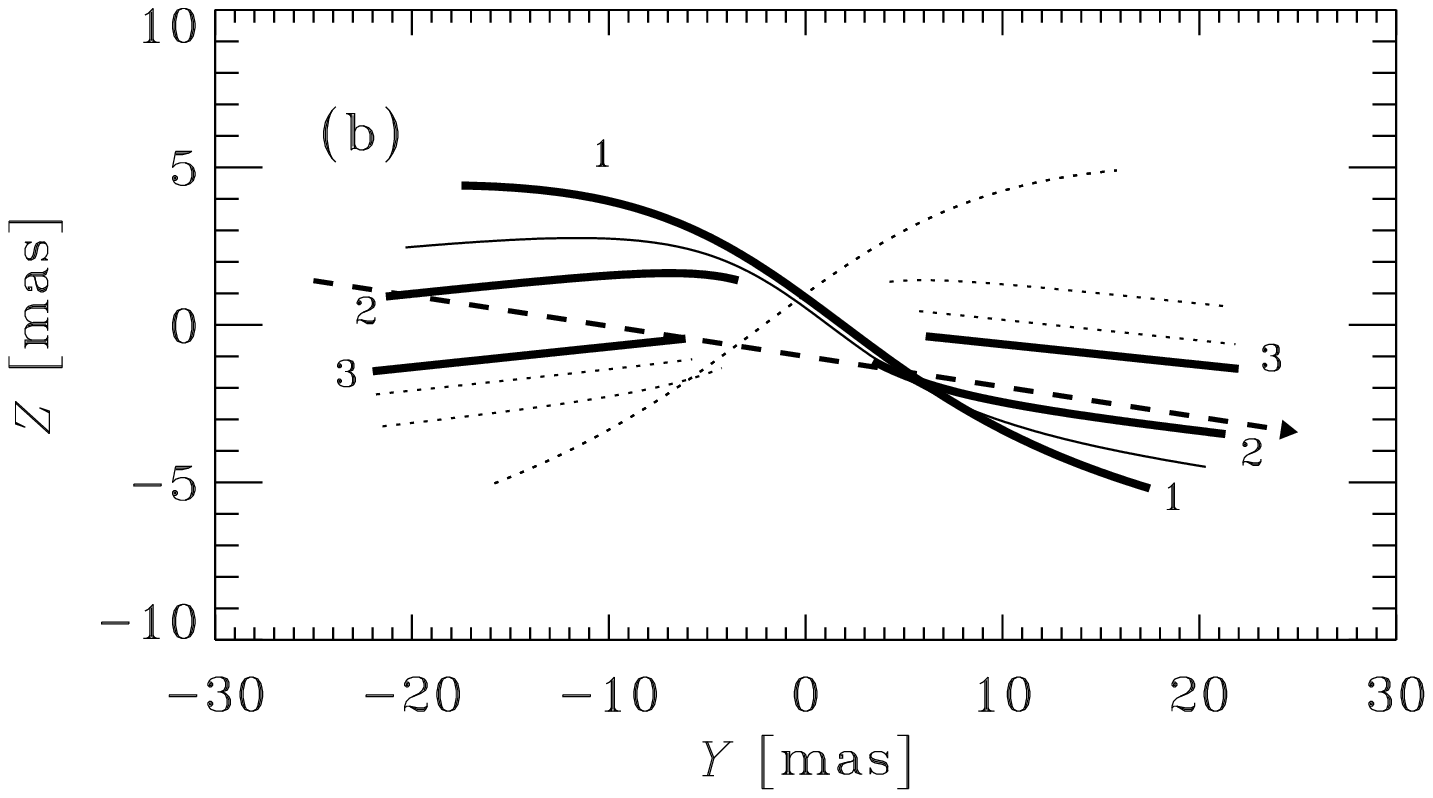}
\end{center}
 \caption
{(a) Three-dimensional image of the mid-plane of the masing disc
with $m=2$ and of the maser spots in V778 Cyg. The line of sight
lies in the $YZ$ plane and makes an angle of $5.5^{\circ}$ with
the $Y$-axis. The coordinates of the disc centre are (0,0,0).
The grey (green) spots show positions of masers. \protect \\
(b) Cross-sections of the disc mid-plane for
$X=-13, -8, -5, 0, 2, 4, 15$ mas. Solid and dotted curves
correspond to negative and positive values of
$X$. The strongest masers at $-22$, $-17$ and $-15$ km s$^{-1}$
appearing at $X\approx -13, -5$ and 0 are
shown as spots in panel (a) and marked by 1, 2 and 3, respectively.
The disc cross-sections are these $X$ are shown by  thick solid curves in
panel (b).
The dashed line with an arrow shows the line of sight.}
\label{fig:3D}
\end{figure*}

Fig.~\ref{fig:tau_lv} shows the dependence of the optical depth in
a masing line $\tau_{\rm m}$ on the impact parameter $b$ and on
the maser velocity $V$ (the figure is identical if we substitute
$v'\rightarrow -V', b'\rightarrow -b'$).  Panel (b) shows that
$\tau_{\rm m}$ as a function of velocity has two peaks: near zero
(systemic) velocity,  where the impact parameter $b\sim 0$, and
near the Keplerian velocity at the outer disc radius $V'\approx
\xi$. In the latter case the dominant rays emerge at an impact
parameter $b\approx 0.9 \Rout$ (for $\xi=11.7$, which is
considered here for application to V778 Cyg).
In general, for a
fixed $b$, the velocity spectrum has two maxima: one corresponds
to almost solid-body rotation with $V(b)\approx V_{\rm K} (\Rout)
b/\Rout$, another corresponds to the Keplerian rotation with $V(b)
\approx V_{\rm K}(b)$.  The first one is much stronger at small
impact parameters, while the latter dominates at larger $b$.

For a given velocity $V$, the optical depth as a function of the
impact parameter $b$ has a single maximum, as in
Fig.~\ref{fig:tau_lv}(a).  If emission at any given velocity is
only detected at a single spatial position \citep[e.g.][]{SS06},
$V(b)$ can be evaluated by fixing $V$ and finding the value of $b$
for which $\tau_{\rm m}$ reaches the maximum, i.e. by solving the
equation $\partial \tau_{\rm m}/\partial b=0$. This gives the
relationship shown by the solid curve in Fig.~\ref{fig:vel}.
It has a solid-body rotation part at small $b$ and a Keplerian
part at large $b$. We also note that $V(b)$ is almost independent
of $\xi$ and $\Rin/\Rout$.

\subsection{Disc geometry}
\label{sec:geometry}

The longest maser amplification length is reached when radiation
passes through the main body of the disc or through its outer
parts. The strongest spectral line for Keplerian discs (with small
ratio of inner-to-outer radii) observed edge-on appears at the
systemic velocity and its spatial position coincides with the
projection of the disc centre.  We therefore associate the
strongest component in the observed maser spectrum of V778 Cyg
(chosen as the origin of coordinates in Fig.~\ref{fig:geom}) with
the systemic component coinciding with the projection of the disc
centre and fix the systemic velocity $V_{\rm sys}$ at
$-15$~km~s$^{-1}$, which is consistent with a value of
$V_{\rm sys}$ obtained from the observations of OH masers
($-15.7$ km s$^{-1}$, \citealt{LM88};  $-15.1$ km s$^{-1}$,
\citealt{BK91}). We also expect to see strong maser features near
the projection of the outer radius of the masing disc at
velocities shifted from $V_{\rm sys}$ by the Keplerian velocity at
this radius.

\begin{table}
\begin{center}
\caption{Model parameters}
\begin{tabular}{|ccccccc|}
\hline
$\Rout$$^a$ &  $i^{\rm \ b}$ & $\Phi^{\ c}$ & $\Theta^{\ d}$ &  $A^{\ e}$  & $V_{\rm sys}$$^f$  &  $M_{\rm s}$$^g$ \\
AU  &  deg & deg    &  deg     &       & km s$^{-1}$    & M$_{\odot}$ \\
\hline $40 \pm 3$ &  $84.5 \pm 1$  & $-8\pm 2$   & $4 \pm 2$ &
$0.23\pm 0.01$ &
  $-$15         & $1.7 \pm 0.1$ \\
  \hline
\end{tabular}
\label{tab:maser}
\end{center}
{ $^{a}$  The outer disc radius.  $^{b}$  The inclination angle of the disc plane.
$^{c}$ The phase angle  (see eq. [\ref{eq:z}]).
$^{d}$  The position angle of the line of nodes.
$^{e}$ The amplitude of the shear wave (see eq. [\ref{eq:z}]).
$^{f}$  Systemic velocity. $^{g}$  The mass of the companion.}
\end{table}

We would like to describe the positions of all maser features
(Fig.~\ref{fig:geom}) and their velocities (Fig.~\ref{fig:vel})
simultaneously. We assume that the masers are produced in a thin
curved disc of a shape described by the following expression
\be \label{eq:z}
z(R, \phi)=A(R)\ R\ \sin [m (\phi - \Phi)],
\ee
where $R$ is the cylindrical radius, $\phi$ is the azimuthal angle and
$z$ is the deviation from the plane $XY$, where the $Z$-axis is
chosen in the direction of the disc angular velocity. $A$ is the
amplitude of the shear wave (which we assume to be constant), and
$m$ and $\Phi$ are its azimuthal number and phase, respectively.
The projection of the disc on the sky depends on the inclination
angle $i$ (the angle between the line of sight and the $Z$-axis)
and on the position angle of the line of nodes~$\Theta$.

Theoretically, it is rather easy to produce a warp with azimuthal
number $m=1$ in a disc in a binary system,  using tidal effects.
The theory of such discs is quite well developed \citep[see, for
example,][and references therein]{PT95}. In that case, rings at
different radii can have different inclination angles and the
observed spatial distribution in Fig.~\ref{fig:geom} could be
reproduced by two such rings: one ring of a larger radius
reproduces maser features at velocities $V \gtrsim -15.9$ km
s$^{-1}$ and  $V \lesssim -18.5$ km s$^{-1}$, while another ring
of a smaller radius produces features at $-18.3$ km s$^{-1}$
$\lesssim V  \lesssim -16$ km s$^{-1}$. However, the observed
linear increase of maser velocities  with the impact parameter
means that all maser features lie on very close orbits. This makes
the $m=1$ warp interpretation improbable.

On the other hand, we find that the observed  $\cal{S}$-like  structure
can be well described by a single orbit with azimuthal mode $m=2$.
We discuss possible reasons for origin of such warp in Section \ref{sec:origin}.
We fit the spatial distribution of maser features and obtain
the set of parameters which are given in Table 1 (parameters $R=\Rout$,
$i, \Phi, \Theta$ and $A$).
For illustration we present  a three-dimensional image
of the disc taking $\Rout=R=40$ AU, assuming that $A$ is constant and $\Rin=10$ AU,
in Fig.~\ref{fig:3D}(a).
 At large impact parameters the coherent length for maser
amplification is large in case of the flat disc. For the warped
disc shown in Figure~\ref{fig:3D}, the line of sight passes
through a large portion of the disc at large negative  values of
$X$, and the maser is bright. However, at large positive $X$ (see
the extreme dotted curve in Fig.~\ref{fig:3D}b corresponding to
$X=15$ mas) it crosses the disc plane at a steep angle reducing
thus the amplification length. At small negative $X$ (see the
thick solid curve 2), the line of sight passes through the whole
disc, while at small positive $X$ (dotted curves) it passes though
one part of the disc only.  This would explain why we only detect
masers from one side of the disc.  We explain this more fully in
the next Section.

\subsection{Maser velocity distribution}

In a doubly-warped disc with $m=2$, the velocities are not precisely
described by the Keplerian rotation law. However,
because the warp is not very large, we neglect possible
deviations from  Keplerian motion.
To describe the velocity distribution of maser features,
we fit the central mass $M_{\rm s}$, while
keeping all other parameters fixed at the
values deduced from the geometry in Sect.~\ref{sec:geometry}.

In Section \ref{sec:geometry}, we showed that the observed dependence of the
velocity on the impact parameter can be well described by a
theoretical $V(b)$ dependence for a Keplerian disc with $M_{\rm
s}=1.7~\msun$ and other parameters from Table 1 (see solid curve
in Fig.~\ref{fig:vel}).  In our interpretation the strongest maser
feature at $-15$~km~s$^{-1}$ coincides with the systemic velocity
(gas moving perpendicularly to the line of sight) and is projected
towards the disc centre $b=0$. The features at
$\approx-17$~km~s$^{-1}$ (i.e. $-2$~km~s$^{-1}$ on
Fig.~\ref{fig:vel}) corresponding to the impact parameter $b\simeq
0.2 \Rout$ show a linear dependence close to $V(b)\approx V_{\rm
K}(\Rout) b/\Rout$ (shown by the dotted line).  If the disc were
flat, these features would be weaker than the systemic feature at
$b=0$, because the optical depth at this position is about 40 per
cent of that at $b=0$ for the assumed gas temperature $T=300$~K
(see Fig.~\ref{fig:tau_lv}).  However, because of the warping, the
line of sight at $b\simeq 0.2 \Rout$ passes almost through the
whole disc (see curve 2 on Fig.~\ref{fig:3D}(b)), whereas the
amplification at $b=0$ takes place through only  half of the disc
 (again because of the warp).
This results in a strong increase of the maser power at these
intermediate velocities and impact parameters.

 The weakest detected component at $-22$~km~s$^{-1}$ (i.e.
$-7$~km~s$^{-1}$ in Fig.~\ref{fig:vel}) lies exactly at the
velocity-impact parameter relationship computed for a Keplerian
disc, which at large $b$ follows a simple  Keplerian rotation law
$V(b)\propto 1/\sqrt{b}$ (the dashed  curve in
Fig.~\ref{fig:vel}). The position and velocity of this component
are consistent with the expected theoretical peak of the maser
emission at $b\approx 0.9\Rout$ and $V\approx V_{\rm K}(\Rout)$
(see Fig.~\ref{fig:tau_lv}).

We now note that for a flat, symmetric disc both blue-shifted and
red-shifted components should be observed around the systemic
velocity.  However, only blue-shifted features are visible
(Fig.~\ref{fig:geom}).  In a warped disc the line of sight cuts
the near and far sides of the disc at different angles. The
cross-sections of the central disc plane are shown in
Fig.~\ref{fig:3D}(b).  The solid and dotted curves correspond to
the projections where the maser amplification would give the blue-
and the red-shifted components, respectively. It is clear from the
discussion in the previous section that the amplification lengths
for the blue-shifted maser components are larger, which would
explain the absence of the red-shifted components in the maser
spectrum.

\subsection{Dust emission}
\label{sec:dust}

\begin{figure}
\centerline{\epsfig{file=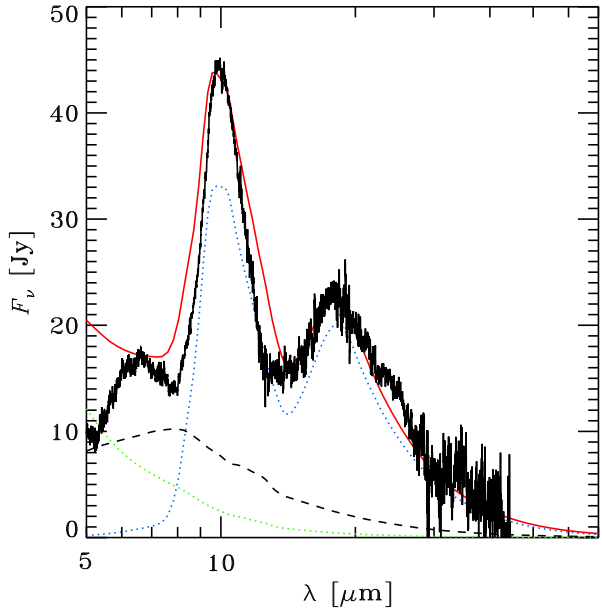,width=8.5cm}}
\caption{The silicate emission bands in the spectra of V778 Cyg
 observed by {\it ISO}/SWS and {\it IRAS}/LRS (Y00).
 The blue dotted and black dashed curves refer to astronomical silicate
and amorphous carbon   model spectra,
respectively.
The dot-dashed curve corresponds to the 2400 K  black body
emission of the carbon star.
The red solid curve shows the total model spectrum.}
\label{fig:dust}
\end{figure}

The infrared spectrum of V778 Cyg obtained by {\it ISO}/SWS and
{\it IRAS}/LRS shows strong silicate features at 10
and 18 $\mum$ (see Fig.~\ref{fig:dust} and Y00). The silicate features as well as the
water maser features are very stable \citep[Y00;][]{EL94}, so it
is natural to assume that the dust and water are co-spatial,
as we can show that similar conditions can explain the observations.
Both can then be locate around the companion in disc, which could have
been formed while presently C-rich star was still O-rich, and as a
result, it contains the O-rich dust and gas which is responsible for the
observed O-bearing dust/molecule features.
However, a small amount of carbon rich dust
from the later stellar wind could also have been captured by the
companion.  We therefore assume that both astronomical silicate
and amorphous carbon grains are present in the masing disc.

We assume that the dust is optically thin and is heated by the
carbon star of radius $R_{*}= 2$ AU and temperature  of
$T_{*}=2400$ K \citep{EL94, BL96}. The heating of the dust by the
companion is negligible if it is a main sequence star.  The
fitting parameters are the dilution factor $W$, the mass of dust,
and the dust grain size $a$. We find that the data are best
described for $a=0.5$~$\mum$. Note, that it is  larger
than the typically assumed grain size of 0.1~$\mum$ in the envelopes of AGB stars
\citep[see e.g.][]{Ju96,Gr06}.
On the other hand, it seems reasonable that grains
would grow whilst in a disc \citep[see][]{LW96,JW01,DC06,DT06},
whereas the smallest grains would be
blown away by the carbon star radiation (see Section
\ref{sec:stability}). The observed IR spectrum is well described
(see Fig.~\ref{fig:dust}) by a model with $W=1.6 \ 10^{-4}$ for
$M_{\rm as}=5.6 \  10^{-7}$ M$_{\odot}$ of astronomical silicate
and $M_{\rm ac}=6.7 \ 10^{-8}$ M$_{\odot}$ of amorphous carbon
(with corresponding temperatures of $T_{\rm as}=290$ K and $T_{\rm
ac}=470$ K obtained from the energy balance). The distance between
the carbon star and the dust material can thus be estimated from
the dilution factor: $D=0.5 W^{-1/2} R_{*} \simeq 80$~AU.

\subsection{Maser pumping model}
\label{sec:maser}

As we argued above, the observed emission of the $6_{16} - 5_{23}$
maser at 22 GHz most probably originates in the disc around the
companion to the carbon star. The hydrogen and water vapor in the disc
are probably mixed with the amorphous carbon and astronomical silicate
dust grains. We assume a solar abundance of elements, giving a
water-to-gas mass ratio of $\fw=6 \ 10^{-3}$ \citep[see, for
example,][]{JW03}. The maser strength depends on the hydrogen number
density $\Nh$, water fraction $\fw$ and the dust-to-gas mass ratio $\fd$
as well as on the gas and dust temperatures.

The dust temperature is determined from  heating by the carbon
star and cooling by its own radiation (see Sect.~\ref{sec:dust}).
The gas temperature is determined by collisions with the dust
grains and radiative cooling by water molecules. We take $\Nh=
10^8$ ${\rm cm}^{-3}$ \citep{KN91} and use $\fd=10^{-2}$
\citep{YD00}, a typical ratio in circumstellar envelopes,
everywhere in the masing disc. We justify our choice of these
parameter values in Sect.~\ref{sec:stability}.  Solving the energy balance
for the gas, we find $T\approx 300$~K at the disc mid-plane.  We
assume the same temperature throughout the disc.

The conditions in the disc are within the range giving rise to
collisionally-radiatively pumped, unsaturated maser emission. In
order to estimate the maser power, we solve simultaneously the
radiative transfer and statistical balance equations for the first
45 rotational levels of the ground and (010) excited vibrational
levels of ortho-$\rm H_2O$. We employ the escape probability
method in order to solve the radiative transfer equation
\citep[for details, see][]{BP04, BP06}.  For a disc half-thickness
of $H\sim10^{14}$~cm at the outer disc edge (obtained by
considering the conditions for hydrostatic equilibrium, see
below), we obtain a maser absorption coefficient of $\kappa=-4.5\
10^{-14}$~cm$^{-1}$.  This gives the maximum maser optical depth
$\tau_{\rm m}\approx 0.73\ \kappa\ \Rout\approx-20$ along the line
of sight at the disc mid-plane in the direction of the maser
component at the systemic velocity (see Fig.~\ref{fig:tau_lv}),
assuming that amplification takes place only on one side of the
disc.  The corresponding maximum brightness temperature is
$T_{\rm b}= -T_{\rm ex} \exp(-\tau_{\rm m}) = 5\ 10^{10} \, {\rm
K}$, for an excitation temperature $T_{\rm ex}=-100$ K.
Averaging the brightness temperature over the disc height
we get $T_{\rm b}\approx 1.5\ 10^{10} \, {\rm K}$. This is above the
lower limit of $6\ 10^8$ K imposed by the observational
data. Thus, the conditions in the disc can easily reproduce the
observed strength of the maser emission.

\section{Discussion}
\label{sec:discussion}

\subsection{Disc stability}
\label{sec:stability}

From the IR spectrum, we have estimated a distance of $\sim 80$~AU
between the carbon star and the region containing the dust.  This
separation is consistent with the scenario in which the dust has been
captured in a disc around a (probably main-sequence) companion
(Y00). Moreover, this estimate is just above the lower limit constrained by the velocity stability of the maser spectral features \citep{SS06}.   We  estimated the radius of the masing disc to be
about 40~AU by fitting the spatial distribution of masers (Table 1). This is
also consistent with the observed velocity profile.

For the disc to be stable, its radius should certainly be smaller than
the distance from the companion to the inner Lagrangian point, which is
about 44 AU, for the the assumed mass of the carbon star of about
$1\msun$ (Y00) and the companion mass of $M_{\rm s}=1.7\msun$.

The  stability of the disc also puts constraints on
 the  dust-to-gas mass ratio  $\fd$. The radiation pressure
$F_{\rm rad}$ from the carbon star acting on the dust should
be balanced by the gravitational force $F_{\rm grav}$ due to the
companion acting on the gas: \be \label{eq:beta} \frac{F_{\rm
rad}}{F_{\rm grav}}=\left( \frac{L_{*}}{4 \pi D^2} \frac{\pi a^2
Q}{c} \right) \left(\frac{G M_{\rm s}}{ R^2} \frac{4 \pi
a^3\rho_{\rm d}}{3 \fd} \right)^{-1} < 1 , \label{eq:c1} \ee where
$Q \simeq 0.2$
 is the extinction efficiency of the dust
 \citep[see][]{OH92} and
 $\rho_d \simeq 3\rm{\,g \,cm^{-3}}$ is the dust density.
For a disc radius of $R=40$ AU, equation (\ref{eq:beta}) gives
$\fd \lesssim 0.02$.

We use the total dust mass in the disc $M_{\rm dust}=M_{\rm
as}+M_{\rm ac}=5.7 \ 10^{-7}$ M$_{\odot}$ to provide an estimate
of the hydrogen number density assuming a homogeneous isothermal
disc: \be \label{eq:nh} \Nh = \frac{M_{\rm dust}}{\fd V_{\rm
disc}m_{\rm H_2}} \gtrsim 5 \ 10^7 \left( {T}/{300\ {\rm K}}
\right)^{-1/2} {\rm cm}^{-3} , \ee where $V_{\rm disc} \simeq 4 H
\Rout ^2$ is the disc volume, $H = \Rout V_{\rm t}/ V_{\rm K} (R)
\simeq 10^{14}$~cm is the disc half-thickness (assuming
hydrostatic equilibrium), $V_{\rm t} \simeq 1.5 ({T}/{300\ {\rm
K}})^{1/2}$ km s$^{-1}$ is the gas thermal velocity, $V_{\rm
K}=6.1$ km s$^{-1}$ is the Keplerian velocity at the outer disc
radius $\Rout=40$ AU and $m_{\rm H_2}$ is the mass of hydrogen
molecule. An additional condition for  disc stability is the
requirement that the radiation pressure must be smaller than the
rate of momentum exchange between the grains and the hydrogen
molecules, giving
another constraint on $\Nh$:
\begin{eqnarray}
F_{\rm rad} \lesssim {m_{\rm H_2} V_{\rm t}}/ {t_{\rm coll}}
\simeq
 m_{\rm H_2} V_{\rm t}^2 \Nh \pi a^2,
\label{eq:c2}
\end{eqnarray}
where $t_{\rm coll}$ is the typical time between  grain-molecule collisions.
We obtain $\Nh \gtrsim 5 \  10^7$ cm$^{-3}$, which is in perfect
agreement with Eq.~(\ref{eq:nh}).
This justifies our choice of  $\Nh=10^8$ ${\rm cm}^{-3}$
and  $\fd=10^{-2}$  when estimating the maser power
in Sect.~\ref{sec:maser}.

\subsection{Origin of a disc with $m=2$}
\label{sec:origin}

We note that at the present time there is  no theoretical
description of how to produce an $m=2$ warped disc \citep[see,
however,][ and reference therein]{LO00}. An $m=2$ warp might be
induced by a non-linear development of warping instabilities due to
the tidal forces from the companion \citep{Lu92, LO00}.  The $m=1$
warping results in density waves described by the azimuthal mode
$m=2$ \citep{PT95}. The effect of  radiation pressure \citep{Pr96}
on these waves can produce an $m=2$ warp.
Such $m=2$ warps cannot realistically have an amplitude $A$ larger
than the disc dimensionless half-thickness $H/R$. Assuming
hydrostatic equilibrium in the disc, we can estimate the ratio
$H/R =V_{\rm t}/ V_{\rm K}(R) \simeq 0.24$, which is of the same
order as the value of $A \approx 0.23$ which we obtain (see Table
1). More theoretical work is required to address the possibilities
of $m=2$ warps  properly.

\section{Conclusions}
\label{sec:conclusion}

MERLIN imaging of water masers from the silicate carbon star V778
Cyg show clear evidence  that they arise from a disc around a
companion to the carbon star.  We estimate the mass of the companion
to be $\sim 1.7$ M$_{\odot}$ and the distance between the stars to be
$\sim80$ AU. This is independently confirmed by the stability
over time of the velocity of the major maser spectral features. The
radius of the masing disc is about $40$ AU. We develop a model
for the disc which is consistent with all available observational constraints.

The silicate dust, observed in the infrared by {\it ISO} and {\it
IRAS}, is likely to be co-spatial with the water vapor producing
the maser emission. This O-rich material must have originated
during an earlier phase of the carbon star and has been captured
in a disc around the companion. The carbon star is about a
thousand times more luminous than its companion and is the major
source of heating in the system.

We find that the observed spectrum of dust radiation is well
described by a combination of $0.5 \mum$ astronomical silicate and
amorphous carbon grains, with masses $M_{\rm as}=5.6 \
10^{-7}\msun$ and $M_{\rm ac}=6.7 \ 10^{-8}\msun$. We model the
conditions necessary for a stable disc, which constrain the
dust-to-gas mass ratio to be $\fd \lesssim 0.02$ and the hydrogen
number density to be $\Nh \gtrsim 5 \  10^7$ cm$^{-3}$. The dust
temperatures of $T_{\rm as}=290$ K and  $T_{\rm ac}=470$ K,
obtained from the energy balance, yield  gas temperature around
300 K (determined by collisions with the dust grains and radiative
cooling by water molecules).  These conditions at the same time
predict very effective water masing and show that the masers could
reach a brightness temperature  more than an order of magnitude larger than the
lower limit, obtained from the observational data.

The observed maser geometry and velocity distribution are
consistent with a doubly-warped Keplerian disc, described by the
azimuthal mode $m=2$. We find that the thickness of the disc and
the warp amplitude of our model are also consistent with the
demands of hydrostatic equilibrium.  We predict successfully the
asymmetric appearance of the disc and explain the origin of the
various maser components.
At present, the theory of such $m=2$ discs
is absent, and this could be an interesting  challenge for
future theoretical studies.

\section*{Acknowledgments}
This work was supported by the Finnish Graduate School for Astronomy and
Space Physics, the Magnus Ehrnrooth Foundation (NB), and
by the grant 2.P03D017.25 of the Polish State Committee for Scientific Research (RS).
We are  grateful to I.~Yamamura for providing the IR spectrum of V778 Cyg
and P.~Ivanov for his help in interpretation of the warped disc.
We thank the referee for useful suggestions.


\label{lastpage}
\end{document}